Astronomic Bioethics: Terraforming X Planetary protection

Authors: Dario Palhares and Íris Almeida dos Santos

Correspondence to: Dario Palhares – dariompm@unb.br

University of Brasília, BRAZIL.

Abstract

A hard difficulty in Astrobiology is the precise definition of what life is. All living beings have a cellular structure, so it is not possible to have a broader concept of life hence the search for extraterrestrial life is restricted to extraterrestrial cells. Earth is an astronomical rarity because it is difficult for a planet to present liquid water on the surface. Two antagonistic bioethical principles arise: planetary protection and terraforming. Planetary protection is based on the fear of interplanetary cross-infection and possible ecological damages caused by alien living beings. Terraforming is the intention of modifying the environmental conditions of the neighbouring planets in such a way that human colonisation would be possible. The synthesis of this antagonism is ecopoiesis, a concept related to the creation of new ecosystems in other planets. Since all the multicellular biodiversity requires oxygen to survive, only extremophile microorganisms could survive in other planets. So, it could be carried out a simulation of a meteorite by taking to other planets portions of the terrestrial permafrost, or ocean or soil, so that if a single species could grow, a new ecosystem would start, as well as a new Natural History. As a conclusion, ecopoiesis should be the bioethical principle to guide practices and research in Astrobiology.

Introduction

Technological advances in space sciences, expanding borders of knowledge and understanding of the universe, have widened our knowledge and deepened philosophical questions such as the origin of life, whether on Earth or on other planets.

Research on extraterrestrial life as a scientific field was consolidated by NASA (National Aeronautics and Space Administration), the US space agency, with the creation of a programme on Astrobiology in 1998[1]. Galante *et al.*[1] consider this policy to be a strong popular appeal for large funding required by this kind of research, that is, Astrobiology, which was created as a scientific and technological discipline by interaction among scientists, society, political and economic groups.

Looking from the Universe, it is possible to highlight some important aspects of the terrestrial ecosystems, especially when human activities have produced unprecedented pollutants, disturbing climatic balance. Astrobiology also has bioethical nuances reflecting on the best for research on the Universe and on some terrestrial practices.

1. Definition of life and delineation of the objectives of research

The first difficulty in the field of Astrobiology is to precisely define what life is. Although the recognition of a life being is simple, intuitive and instinctive , the formal definition of life is very complex[2]. Life beings are distinguishable from the mineral due to their spontaneity and admirable ability of self-replication. They are fragile, perishable and doomed to death, but as they self-replicate, they show that life is strongly resilient: basically, where there is liquid water, life begins.

Science can describe life in operational words, but not in essential words[2]. If there is a definition, life is closer to a verb than to a noun. The autotrophic life beings transform the brute mineral into themselves, that is, they give life to the brute environment. So, it is possible to infer an undefined border between brute minerals and life. Indeed, since 1924, biochemical research Oparin had shown that small simple molecules such as carbon dioxide, ammonia, sulphur, water, etc., and random chemical reactions produce complex molecules such as amino acids, lipids and even polymers that combine in coacerved structures that somehow resemble a living cell[1].

However, despite the countless chemical reactions that occur inside a living cell and the possibility of laboratory synthesis of practically all chemical compounds of a living cell, the vital organization is something transcendent to the mere concrete matter[3]. Living is a metabolic and physiological adjective that describes a living being. Overall, a living being can be in the stage of 'living', of 'dormancy/latency' or 'dead'. A dead living being will never return to life. Some species produce propagules—seeds, spores, buds, etc.—that can remain latent (or dormant) for thousands of years, but when exposed to favourable environment can grow and self-replicate (living stage)[4]. However, the simple preservation of a living being in liquid helium, formalin, conservatives, etc. can maintain the skeletal structure but clearly it is a dead one . Until now, life remains an abstract and an unknown concept, a mysterious force at the same time existing and transcending to a living being[2.3].

All living beings, with no exception, have a cellular structure. So, with absolute ignorance of a living being not organised in cells, it is not possible to have a broader concept of life, and in Astrobiology the search for extraterrestrial life is restricted to extraterrestrial cells. Such epistemic positioning is related to concrete questions. Space exploration must be restricted to mere observation of planets, under the belief that there are life forms that do not belong to cells? Or on the contrary, the cell being so wonderful endeavours must be made for colonisation with terrestrial living beings?

2. Origin of living beings

Fossils show that living beings existed on the terrestrial surface at least for 3.8 billion years[1,5]. Considering that the age of Earth is approximately 4 billion years, it means, in astronomical terms, that as soon as the Earth was formed and contained liquid water on its surface, living beings began appearing. Terrestrial biodiversity is based on the cell with the genetic code of DNA/RNA, whose metabolism is based on the breakage of ATP. This points to what would be the first terrestrial microbe, or LUCA—Last Ultimate Common Ancestor—from which all the terrestrial biosphere originated[1].

This geological aspect brings up deeper questions. If, according to the Theory of Spontaneous Generation, the cell originated on the terrestrial surface from random reactions on mineral substrates, then were there other patterns of cells? If so, what were they? Was LUCA a single event or had it been repeated several times? Were there other genetic systems supplanted by DNA/RNA? Why such primitive patterns are not seen nowadays?

On the last question, Darwin wrote in 1871[1]: "*it is often said that all the conditions for the first production of a living being are now present, which could ever have been present. But if (and oh what a big if) we could conceive in some warm little pools with all sort of ammonia and phosphoric salts—light, heat, electricity present, that a protein compound was chemically formed, ready to undergo still more complex changes, at the present such matter would be instantly devoured, or absorbed, which would not have been the case before living creatures were formed*".

Indeed, the strong relation between liquid water and the presence of living beings is significant. However, no life form is found only in extremely arid deserts; otherwise, wherever there is liquid water there are living beings, including apparently hostile environments. This intrinsic relation guides the astronomical observation of what would be a habitable planet: the one with liquid water on the surface.

In this sense, Earth, the Water Planet, is an astronomical rarity[1,5]. Although water is a common and abundant substance in the Universe, for a planet to show liquid on its surface a certain set of parameters, such as adequate distance from the star, the presence of a magnetic field that protects the atmosphere from the star winds, adequate concentration of glasshouse gases in the atmosphere, etc., are required. Apparently, the space probes sent up now suggest that the planet Mars and the moons Europa and Ganimedes that orbit Jupiter have liquid water in their subsoil, below an ocean of ice[1]. The moon Europe has an atmosphere rich in oxygen, similar to the terrestrial one, suggesting that liquid water on the surface is undergoing photolysis and releasing the gas. Moreover, the moons of Saturn, Titan and Enceladus, the Uranus moon Titanus and the Neptune moon Triton perhaps have liquid water under ice in the subsoil[1].

Since the ancient Greece, the origin of terrestrial life has been imagined as the result of a panspermia—that the Universe has living beings that are transported somehow from one planet to another[6]. Anaxagoras mentioned this idea as early as 500 BC, and this concept was recirculated at the end of the 14th century by scientists devoted to organic chemistry and biochemistry[6]. In 1908, the physicist Arrhenius published a book in which he discussed the Theory of Panspermia[7]. According to it, the living beings are one of the elements of the Universe, who would be transferred from one galaxy to another via comets and asteroids until they reach a favourable planet to grow. Indeed, recently, some rocks were found on the Earth, which originated from Mars and had traces of ancestral organic matter[5,8].

The stronger criticism against the Theory of Panspermia is that it does not clarify about the exact mode of origin of the living cell. However, it has deep bioethical impact on it, if panspermia is a phenomenon intrinsic to the Universe, the living beings of each planet would have somehow gone to the neighbouring planets, and in the case of the Earth humans are the only species with the ability spread living beings outside Earth.

3. Planetary protection

Astronomer Huygens had described channels on the surface of Mars as early as 1659[1]. Arrhenius[7] considered it plausible that there might be life on Mars, which however has not yet been confirmed. With effectively conquering space by the 1960s, with astronauts going to the orbit, it appeared as concern the physiological effects of microgravity and, in parallel, a fear of contraction of infectious diseases spreading from space and/or other planets that could cause massive human mortality[9]. Fearing cross-infection, and on the bioethical principle of planetary protection, the study and observation of neighbouring planets would be via sterilized equipment to avoid any interplanetary contamination. Indeed, studies on space have been confined to the study of the planets with probes and robots with no reports of intentional colonisation of neighbouring planets by terrestrial living beings.

In the interests of planetary protection, it is important to highlight some aspects of the terrestrial Natural History. About 1.5 billion years after the appearance of the first living cells, the first photosynthesising living beings appeared and, after catalysing the photolysis of water, strongly enriched the terrestrial atmosphere with oxygen[1]. So, about 2.1 billion years ago, the first multicellular living beings appeared. Nowadays, all the multicellular megabiodiversity breathes and obtains energy from controlled combustion of organic compounds[1]. All the species that do not require oxygen are unicellular[1].

Oxygen is a highly reactive gas, so the living beings evolved to biochemically control this dimension[10]. In comparison, sulphur belonging to the same family in the periodic table is also highly reactive, but is also highly toxic: in human beings, its inhalation at a concentration of 0.001 ppm is irritant to the respiratory mucosa, while it is lethal at a concentration of 0.01 ppm[11]. The atmosphere of the moon Io of Jupiter is highly rich in sulphur, and any terrestrial multicellular organism that went inside that moon would immediately be corroded by that strong oxidant[1]. Hypothetically, if a living being from Io went to Earth, it would be quickly oxidised by the corrosive oxygen.

So, it is clear that the eventual biological colonisation of the neighbouring planets would be feasible only with microbes adapted to the environmental conditions in the colonising planets. So, one of the sub-disciplines of Astrobiology is the study of the so-called extremophile microbes, that is, microbes adapted to some terrestrial environments whose physico-chemical characteristics are similar to other planets[5]. The principal categories of extremophiles are: thermophilic and hyperthermophilic (adaptation to extreme heat, close to the boiling point of the water); psychrophylic (adaptation to cold), acidophilic, alkaliphilic, barophilic (adaptation to acid, alkali and high pressure) and halophilic (adaptation to high saline concentrations)[1]. Some of these groups are of particular interest like the ecosystems inhabiting the gabbroic rocks from 1,400 m below the ocean floor and the polar ecosystems living at a freezing temperature[1,8].

Beside extremophile microbes, the chemiolitotrophic microbes are also of interest. Such microbes are autotrophic and obtain energy from the oxidation of inorganic compounds such as phosphorus, sulphur, ammonia, iron, arsenic, selenium, etc. That is, the entire ecosystems exist in the absence of light. The case of the archea *Desulforudis audaxvitor* is intriguing because it grows in dark mines rich in uranium and obtains energy from the radiolysis of water from the decaying uranium[1].

Anyway, the fear of an eventual interplanetary colonisation with terrestrial microbes is related to the ecological field of invasive species, a phenomenon that is almost totally related to human travels and intentional or occasional transport of living beings from one ecosystem to another. Downey and Richardson[12] listed essential aspects of the ecology of invasive species: data show that ecosystems related to restricted regions (islands, lakes, pools, etc.) are more vulnerable to invasive species, that the documented extinctions were caused by predator animals and the extinction of native plants related solely to the introduction of an invasive plant has not been observed, since the invasive plants often occupy places already disturbed by human activities. Anyway, the ecological inter-relationship among native and invasive plants is very complex and the extinction of native plants would take a long period, perhaps hundreds of years[12].

On the ecological relationship among microbes, Veresoglu *et al.*[13] highlight that all living species are doomed to extinction as we know them, either by disappearance or origin of new species. The microbes that lived on the Earth 3.8 billion years ago don´t exist anymore, although they are responsible for the origin of all living species today. Anyway, the microbial ecosystems have species with great ability to metabolic adaptation, such as microbes that can live in external environments but sometimes live with roots of plants or the guts of animals.

On the introduction of new microbes into an ecosystem, Veresoglu *et al.*[13] say that the microbial ecosystems are very complex and are organised in layers such that the simple presence of a new species would be insufficient to significantly impact them. Instead, the human modifications to the environment (asphalt, drainages, etc.) completely change the characteristics of the environment, and would strongly change the microbial ecosystems.

So, the group of ecological evidences points to an eventual colonisation of planets with terrestrial living beings that would be feasible only with extremofile microbes, which has little potential of impacting an ecosystem if one is already present in those planets.

Terraforming

The word terraforming appeared first in science fiction literature in 1942 in the magazine *Astounding Science Fiction* edited by Jack Williamson[14]. From a bioethical position opposed to planetary protection, terraforming is the intention of modifying the environmental conditions of the neighbouring planets, particularly Mars, in such a way that human colonisation would be possible.

Alexandrov[9] defends the bioethical principle of terraforming. He has identified the difference between anthropocentrism and biocentrism. Anthropocentrism attaches great value intrinsic to humanity, while biocentrism puts humans as a natural enemy of all the biosphere, including the extraterrestrial biosphere, if there was one. So, to Alexandrov[9], if planetary engineering could be carried out to benefit humanity and create a human colony, then such effort should be made. He believes that if terraforming is feasible, somebody will do it sooner or later.

The theoretical steps for eventual terraforming aslisted by Haynes and McCay[15] are: first, a prospection of the surface of Mars; second, colonisation with terrestrial microbes; third, eventual environmental changes with engineering acitivities, for example, with nuclear explosions; fourth, the cultivation of food plants and finally human colonisation. The authors consider that microbial colonisation is relatively plausible with the existing technology, but human colonisation demands unimaginable technological resources[16].

In Earth, the microbes are the beginning and the end of the ecological cycle. The ecological system is cyclical because the wastes of a given living being are substrate to other species, principally microbes. So, human pollution derives from the production of organic waste, principally plastics that, although chemically burnable, are not naturally degradable by any microbial species[2]. The ideal terraforming would be by microbial species that colonised the other planets and could change the environment, such as has occurred to Earth when the first photosynthesising living beings appeared[15].

So, one of the research fields in Astrobiology is to identify and select extremophile microbes and carry out experiments of resistance to space environment and survival in extraterrestrial environments. Tarashashvili and Aleksidze[8] created in laboratory a growing medium similar to what the Martian soil would be, then they inoculated samples of extreme environments containing iron-bacteria, silicon-bacteria, sulphur-bacteria, mycobacteria and cyanophytes that hadn´t been identified yet. It was possible to see the growth of colonies of those microbes, so the authors suggest cultivation and amelioration to obtain strains considered as promising to colonise Mars.

Horneck *et al.*[4] studied spores of the bacteria *Bacillus subtilis* in experiments carried out on satellites of ESA (European Space Agency) and showed that the spores were highly resistant to vaccum, to freezing temperatures and to radiation from Space. Although the spores were highly sensitive to ultraviolet light, a thin layer of rock was enough to protect them. Those experiments suggest that spores of this bacterial species could survive for an undefined period, perhaps thousands of years, corroborating the Theory of Panspermia and show that it would be possible to transport viable microbial spores from Earth to other planets.

Friedmann and Ocampo-Friedmann[17] suggested that *Cyanophycea chroococcidiopsis* would be an ideal candidate for transplantation to other planets, since it is ubiquitous on Earth, and can be considered as a microbial living fossil. However, Thomas *et al.*[18] simulated in laboratory the climatic conditions of Mars and tried growing Chroococcidiopsis and other photosynthesising microbes, but they did not survive.

Paulino-Lima *et al.*[19] studied the archea *Deinococcus radiodurans* and documented its resistance to radiation and ultraviolet rays, but since this is a heterotrophic aerobic microbe, it has a lesser chance of being a probable candidate to be a pioneer in extraterrestrial colonisation.

Final consideration

The two bioethical positions—terraforming and planetary protection—are antagonistic. Space exploration needs large funding and support and up to now just a few worldwide agencies have invested on this field, and in a paradigm that astronomic biology is concerned with planetary protection. Terraforming is still at a stage of hypothesis and bioethical debate but till date, no experiments have been carried out to take microbes purposely to other planets.

The fear of space agencies of damage to other planets with terrestrial living beings is not supported by ecological studies of the terrestrial ecosystems. It is clear that an eventual interplanetary colonisation could only be feasible by extremophile microbes, which, in case of matching living beings already existing, would be of great competitive disadvantage. But if those microbes found even minimally favourable conditions, they would create a new Natural History on that planet.[4]

Notwithstanding, if anyone of our neighbouring planets has life, or at least life of the same duration as of the Earth (around 3.8 billion years), the living beings would have already developed a cover in these planets, occupying earth, depths, atmosphere. After all, life is one of the wonders of the Universe, and when established on a planet, will evolve in such a way as to occupy the maximally possible area of the planet[2]. The living being is born and dies, but after each replicating cycle, the living biomass of the planet increases continuously. However, our neighbouring planets do not have any form of life, either a cell or an unimaginable structure.

From this bioethical antagonism, a third concept that synthesises the extremities, ecopoiesis, has emerged. Ecopoiesis is a word mentioned by Robert Haynes in 1984 and published in 1989[20] and is related to the creation of new ecosystems in other planets. This terminology summarises the former concept of planetary ecosynthesis, first coined in 1979 by Averner and MacElroy[21].

That is, neither a radical transformation to allow human colonisation nor the omission facing a scientific-technological ability. If a living being comes only from another living being, then life shows to be a force, something beyond our reach, present in the Universe, capable of transforming the inorganic into organic and, in the evolutionary fight for survival, transform a whole planet in an ecosystem each time deeper and more complex[22].

The theories of origin of life—Spontaneous Generation and Panspermia—are not mutually excluding, and can be complementary. Under a cosmogonic perspective, since in a planet some kind of living beings appear, they somehow colonise the other planets, either by random means, such as collision with asteroids, or by means of intelligent forms that do that intentionally[5,15,16]. Moreover, if the Theory of Panspermia is correct, it is even possible that extraterrestrial microbes hit Earth with some frequency, but these life forms would face strong difficulties to resist to the toxicity of oxygen, to the stressful climate, but, foremost, to compete with other life forms already established here.

Therefore, ecopoiesis is related to the bioethical position of cosmocentrism, that is, planets that house some form of life have an intrinsic value higher than the one with only mineral elements[5,15,16]. So, Mautner and Matloff[16] and their Society for Life in Space defend that ecopoiesis is the cosmogonic proposal of human existence, that is, one of the tasks to Humanity would be one of disseminating terrestrial propagules to the neighbouring planets, as it is the only species able to do that.

The idea of ecopoiesis can also guide future expedition to other planets: one line of research can be the complex task of identifying, isolating, cultivating and preserving 'the' microbe candidate to colonise another planet. But if Panspermia occurs as a meteorite coming from a planet, what reaches the planets is not a single species, but a whole ecosystem preserved in rocks. So, for example, the colonisation of the moon Europa of Jupiter could be carried out by leaving there, or perhaps burying there, a portion (some kilograms, half a ton, or so ) of the Artic or Antarctic permafrost. Or, samples of the oceans or of the terrestrial soils could also be transported. That is, if a whole microbial ecosystem is transported similar to what would be a meteor to the neighbouring planet, it would be enough if a single species grew to start a new ecosystem.

In conclusion, ecopoieses is based on bioethical principles that should guide space exploration in substitution to the principle of planetary protection.